\documentclass[%
 reprint,
 amsmath,amssymb,
 prl,
]{revtex4-2}
\usepackage{graphicx}% Include figure files
\usepackage{dcolumn}% Align table columns on decimal point
\usepackage{bm}% bold math
\usepackage{amsmath}
\usepackage[hidelinks]{hyperref}% add hypertext capabilities
\usepackage{cleveref}
\usepackage{natbib}
\usepackage{aas_macros}
\begin{document}

\preprint{APS/123-QED}

\title{Predicting spatial curvature \(\Omega_K\) in globally $CPT$-symmetric universes}% Force line breaks with \\
%\thanks{}%

\author{Wei-Ning Deng}
\email{wnd22@cam.ac.uk}
% \affiliation{Astrophysics Group, Cavendish Laboratory, J.J. Thomson Avenue, Cambridge, CB3 0HE, UK}
% \affiliation{Kavli Institute for Cosmology, Madingley Road, Cambridge, CB3 0HA, UK} 
% \affiliation{Queens College, Silver Street, Cambridge, CB3 9ET, UK}

\author{Will Handley}%
\email{wh260@cam.ac.uk}
\affiliation{Astrophysics Group, Cavendish Laboratory, J.J. Thomson Avenue, Cambridge, CB3 0HE, UK}
\affiliation{Kavli Institute for Cosmology, Madingley Road, Cambridge, CB3 0HA, UK} 
% \affiliation{
%  Gonville \& Caius College, Trinity Street, Cambridge, CB2 1TA, UK
%  }

%\collaboration{}%\noaffiliation

\date{\today}% It is always \today, today,

\begin{abstract}
    Boyle and Turok's $CPT$-symmetric universe model posits that the universe was symmetric at the Big Bang, addressing numerous problems in both cosmology and the Standard Model of particle physics. We extend this model by incorporating the symmetric conditions at the end of the Universe, which impose constraints on the allowed perturbation modes. These constrained modes conflict with the integer wave vectors required by the global spatial geometry in a closed universe. To resolve this conflict, only specific values of curvature are permissible, and in particular the curvature density is constrained to be $\Omega_K \in \{-0.014, -0.009, -0.003, \ldots\}$, consistent with \textit{Planck} observations.
\end{abstract}

\maketitle

\section{Introduction}
The concordance $\Lambda$CDM model and the Standard Model successfully explain a range of observed and theoretical phenomena in cosmology~\cite{2020A&A...641A...6P} and particle physics~\cite{1995iqft.book.....P}. However, several issues remain unresolved: For cosmology, there are the horizon~\cite{1972grun.book.....D} and flatness~\cite{1983Natur.305..673C} problems, as well as the composition and nature of dark matter and energy. For the standard model of particle physics, there is the cosmological constant problem~\cite{1989RvMP...61....1W}, Weyl anomaly~\cite{1994CQGra..11.1387D} and the strong $CP$ problem~\cite{1976PhRvL..37....8T}. Attempts to address these challenges often involve introducing new degrees of freedom, such as an era of inflation~\cite{1981PhRvD..23..347G}, a wide variety of dark matter particles~\cite{2019Univ....5..213P}, axions~\cite{1977PhRvL..38.1440P}, or modifying General Relativity~\cite{2012PhR...513....1C}. Despite these efforts, no single model can resolve all these issues simultaneously, and such modifications compromise the apparent simplicity of our universe as suggested by recent observations~\cite{2020A&A...641A...6P, 2024ApJ...966..157F}.

To address these challenges, \citet{2018PhRvL.121y1301B, 2022AnPhy.43868767B} propose a novel cosmological model known as the $CPT$-symmetric Universe. Rather than introducing new degrees of freedom, they posit that the universe is $CPT$-symmetric at the Big Bang singularity. Given this initial condition, the aforementioned issues could be solved without adding extra complexity~\cite{2021arXiv211006258B,2022arXiv220810396B,2024PhLB..84938442B,2023arXiv230200344T}. For example, the $CPT$-symmetric universe exhibits a periodic global structure, which allows for the calculation of its gravitational entropy~\cite{1977PhRvD..15.2752G}. As noted by \citet{2024PhLB..84938443T}, maximizing this entropy resolves the flatness, homogeneity, and isotropy problems without requiring inflation.

In this letter, we further investigate the impact of this periodic global structure on cosmological perturbations. Previous work by \citet{2022PhRvD.105h3514L} has explored this concept within the context of flat universes. Their findings suggest that the periodic conditions impose constraints on perturbations within the universe, which have observable effects on the Cosmic Microwave Background (CMB) power spectrum~\cite{2022PhRvD.105h3515B,2022PhRvD.105l3508P}.

Building on this work, we extend the analysis of \citet{2022PhRvD.105h3514L} to include curved universes. Our findings indicate that the constraints imposed by the future conformal boundary conflict with the spatial boundary conditions in a closed universe. To reconcile this conflict, only specific values of curvature are permissible, which are consistent with the observations from \texttt{Planck 2018}~\cite{2020A&A...641A...6P}, adding to the recent conversation in the literature surrounding closed universes~\cite{2020NatAs...4..196D, 2021PhRvD.103d1301H, 2022MNRAS.517.3087G, 2020MNRAS.496L..91E}.

\section{Background and perturbations} 
We build on the derivation presented in \citet{2022PhRvD.105h3514L} and extend it to a curved universe. The perturbed FLRW metric can be written as:
\begin{equation}
    ds^2 = a^2 \left[-(1 + 2\Phi)d\eta^2 + (1 - 2\Phi)\left(\frac{1}{1-\kappa r^2}dr^2+r^2d\Omega^2\right) \right],
\end{equation}
where \(\Phi\) is the perturbation in the Newtonian gauge, and $\kappa$ controls the spatial curvature of the universe.

We analyse the evolution of the background in conformal time \(\eta\) instead of physical time to render explicit the periodic behaviour of the universe. Following Lasenby~\cite{2022PhRvD.105h3514L}, only three components are considered: dark energy $\lambda$, radiation $r$, and spatial curvature $\kappa$, omitting matter. The zero-order Friedmann equations are: 
\begin{equation}\label{eq:friedmann_eq}
    \dot{a}^2 = \tfrac{1}{3}\lambda a^4 - \kappa a^2  + \tfrac{1}{3} r, \qquad 
    \ddot{a} = \tfrac{2}{3}\lambda a^3 - \kappa a,
\end{equation}
where \(\hbar = c = 8\pi G = 1\) and dots denote \(d/d\eta\).

\begin{figure*}[!ht]\centering
     \includegraphics{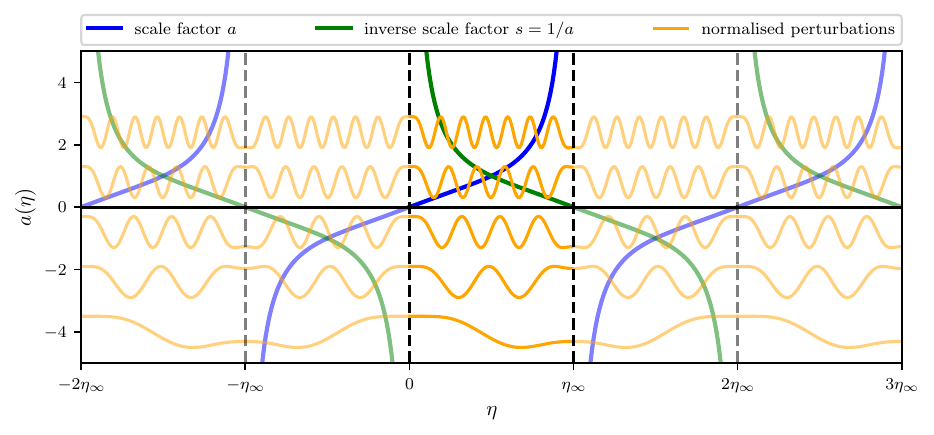}
     \caption{Evolution of the scale factor \(a\), its reciprocal \(s\equiv 1/a\), and the first five allowed perturbations which are (anti)-symmetric about the future conformal boundary. The perturbations have been normalised to have constant oscillatory amplitude across cosmic time for clarity.}
     \label{fig:a_perturbations-eta}
\end{figure*}

The solution to \cref{eq:friedmann_eq} can be expressed as a Jacobi elliptic function~\cite{2021arXiv210906204B}, which exhibits periodic global structure in conformal time. The universe begins at the Big Bang \(a = 0\), evolves to the end of the universe at \(a=\infty\) at a finite conformal time \(\eta_{\infty}\) termed the future conformal boundary (FCB), and then returns from the future boundary back to a future Big Bang, starting the cycle over again. However, this cyclic behaviour does not by default hold for perturbations. To maintain the behaviour, the perturbations should be symmetric not just at the Bang, but also at the FCB. The symmetry constrains the allowed modes of perturbations (see \cref{fig:a_perturbations-eta}).

To analyse the perturbations, we write the first-order Einstein equation as:
\begin{equation}\label{eq:perturbation_eq}
    \ddot{\Phi} + 4aH\dot{\Phi} + \tfrac{1}{3}(k^2 + 4a^2\lambda-12\kappa)\Phi = 0.
\end{equation}
where \(k\) is the comoving wave vector of the perturbation. Switching the independent variable from conformal time $\eta$ to scale factor $a$, one finds
\begin{equation}\label{eq:perturbation_a}
    (\lambda a^2 - 3\kappa + \frac{r}{a^2} )\Phi'' + (6\lambda a - \frac{15\kappa}{a} + \frac{4r}{a^3})\Phi' + (4\lambda+\frac{k^2-12\kappa}{a^2})\Phi = 0,
\end{equation}
where a prime denotes \(d/da\).

One degree of freedom between \(r\) and \(\lambda\) can be removed from \cref{eq:friedmann_eq,eq:perturbation_a} by fixing the scaling of \(a\). Following~\cite{2022PhRvD.105h3514L}, \(r = \lambda\) is chosen, equivalent to demanding \(a=1\) when the energy density of matter radiation and dark energy are equal, after which \cref{eq:perturbation_a} becomes
\begin{equation}\label{eq:Phi_ODE}
    a(a^4 - 2a^2 \tilde\kappa + 1) \Phi'' + (6a^4- 10a^2 \tilde\kappa + 4 ) \Phi' + a(4a^2 + \tilde{k}^2-8\tilde\kappa) \Phi = 0, \nonumber
\end{equation}
\vspace{-2.5em}
\[ \text{where} \qquad
    \tilde{k} \equiv k/\sqrt{\lambda}, \qquad \tilde \kappa \equiv \frac{3}{2\lambda}\kappa\]
are the dimensionless wave vector and curvature respectively. This equation reduces to eq.~(23) in~\citet{2022PhRvD.105h3514L} when \( \tilde\kappa = 0 \).
Note that if we replace \(\Phi(a)\) with \(\varphi(a) \equiv a^2 \Phi(a)\), the equation remains invariant under the transformation \({a\leftrightarrow1/a}\). This symmetry is important in the next section when considering perturbations passing through the FCB as \(a\rightarrow \infty\).

This equation can be solved analytically, yielding two solutions. The solution can be expressed in terms of Heun functions:
\begin{equation}
\begin{aligned}\label{eq:Heun}
    \Phi(a) &= \text{HeunG} \left( -1+2\tilde\kappa \tilde a_+^2,\frac{\tilde k^2-8\tilde\kappa}{-4 \tilde a_-^2},\frac{1}{2},2, \frac{5}{2},\frac{1}{2},a^2\tilde a_+^2\right)\\
    &\tilde a_{\pm}=\sqrt{\tilde\kappa\pm\sqrt{\tilde\kappa^2-1}}
\end{aligned}
\end{equation}
where \(\tilde a_{\pm}\) are the extremum values of the scale factor, corresponding to \(\dot{a}=0\) in \cref{eq:friedmann_eq}. Since the perturbation is frozen at the beginning and starts to oscillate after entering the horizon \(k = aH\), a specific linear combination of these solutions is selected to ensure that \(\Phi(a =0) = 1\) and \(\Phi'(a = 0) = 0\).

\section{Constraints from the future conformal boundary}\label{sec:constrain_FCB}
Next, we investigate the behaviour of perturbations near the FCB.  For universes with curvature \( \tilde{\kappa} \equiv \frac{3}{2\lambda}\kappa > 1 \), the scale factor \(a\) reaches a maximum before they collapse in a ``big crunch'', which are not physically relevant, since we observe an accelerating Universe today. This implies that the Universe cannot possess excessive curvature, establishing an upper limit of \(\tilde{\kappa} < 1\). In this case, the scale factor \(a\) approaches infinity at the FCB.
Consequently, we should consider the reciprocal scale factor \( s \equiv 1/a \). Because of the symmetry in \cref{eq:Phi_ODE}, the solution can be expressed as \(\Phi(s)= s^4 \Phi(a(s)) \). As a result, the solution would continue oscillating through the FCB until the future Big Bang. \citet{2022PhRvD.105h3514L} found that most perturbations diverge before reaching the future Big Bang. Only those solutions that are finite and symmetric or antisymmetric at the FCB remain non-singular (see \cref{fig:a_perturbations-eta}). This symmetry requirement means that only a discrete spectrum of wavevectors are allowed.

To calculate the discrete spectrum of wave vectors, we re-write the perturbation solution into an integral form \cref{eq:integration_sol}, where the sine component captures the solution's oscillatory nature:
\begin{equation}\label{eq:integration_sol}
\begin{aligned}
    \Phi(a) &= \frac{3\sqrt{a^4 + (\tilde{k}^2 -2\tilde\kappa)a^2 + 1}}{a^3\tilde{k}\sqrt{(\tilde{k}^2 -2\tilde\kappa)^2 - 4}} \sin\theta(\tilde k,\tilde\kappa,a),\\
    \theta = \int_0^a &\frac{\tilde{k} a'^2 \sqrt{(\tilde{k}^2 -2\tilde\kappa)^2 - 4}}{\sqrt{1 + a'^4 - 2\tilde\kappa a'^2} \left( a'^4 + (\tilde{k}^2 -2\tilde\kappa)a'^2 + 1 \right)} da',
\end{aligned}
\end{equation}
The integration in \(\theta(a)\) can be expressed as an incomplete elliptic integral of the third kind \(\Pi\):
\begin{equation}\label{eq:psi_elliptic}
\begin{aligned}
    \theta(a) = \tilde{k} \tilde a_- \times \Bigg[ &\Pi\left( \frac{a}{\tilde a_-}, -\frac{1}{2}\tilde a_-^2 \tilde K_{-}^2, \tilde a_-^2 \right) \\
    - &\Pi\left( \frac{a}{\tilde a_-}, -\frac{1}{2}\tilde a_-^2 \tilde K_{+}^2, \tilde a_-^2 \right) \Bigg],\\
\end{aligned}
\end{equation}
\vspace{-1.5em}
\begin{equation*}
\begin{aligned}
    &\Pi(z,\nu,k)=\int_0^z \frac{1}{(1-\nu t^2)\sqrt{1-t^2}\sqrt{1-k^2t^2}} dt,\\
    &\tilde K_{\pm}=\sqrt{(\tilde{k}^2 -2\tilde\kappa) \pm \sqrt{(\tilde{k}^2 -2\tilde\kappa)^2 - 4}}.
\end{aligned}
\end{equation*}
For the flat case (\(\tilde\kappa=0\)), the solution \cref{eq:integration_sol,eq:psi_elliptic} reduces to eq.~(26-28) in~\citet{2022PhRvD.105h3514L}.

\begin{figure}
     \includegraphics{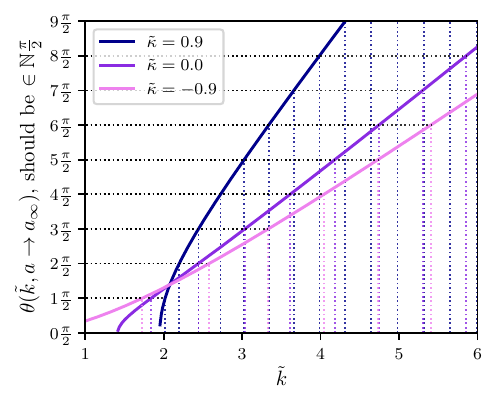}
     \caption{Phase of perturbations \(\theta(\tilde{k},a\rightarrow \infty)\) with different curvature \(\tilde\kappa\). Larger \(\tilde\kappa\) results in a steeper curve. The value of \(\theta\) should be \(\frac{\pi}{2}n, n\in\mathbb{N}\), where the corresponding \(\tilde k\) are the allowed wave vectors forming a discrete spectrum.}
     \label{fig:theta_diffK}
\end{figure}

If we require the solution to be (anti)symmetric at the FCB, the cycles of oscillation should be complete at the boundary. This implies that the angle \(\theta\) in the sine function should equal \( n\frac{\pi}{2} \) at \( a \rightarrow \infty \). 
\begin{equation}\label{eq:theta}
    \theta(\tilde{k}, \tilde\kappa, a \rightarrow \infty) \overset{!}{=} \frac{n\pi}{2}.
\end{equation}
The corresponding \(\tilde{k}\) are the allowed values of discrete wave vectors. The function \(\theta(\tilde k,a\rightarrow\infty)\) with different \(\tilde\kappa\) is illustrated in \cref{fig:theta_diffK}. While increasing the curvature value \(\tilde\kappa\), \(\theta\) becomes steeper, and the corresponding discrete wave vectors \(\tilde{k}\) have smaller spacing. 

\section{Implications for spatial curvature \(\Omega_K\)} 
These observations have significant implications for closed universes, where the compactness of spacial slices necessitates integer wave vectors (Sec.~2.4 in~\cite{2017CCoPh..22..852T}). This requirement is generally in conflict with the discrete spectrum of wave vectors arising from FCB considerations, prompting the question of whether this conflict can be resolved by matching the two discrete sets.

\begin{figure}
     \includegraphics{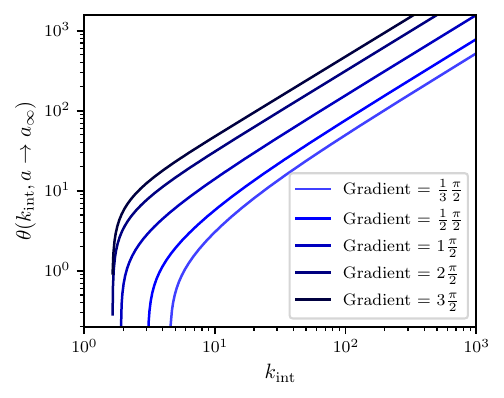}
     \caption{Phase of perturbations $\theta$, in log-log plot. In the high-\(k\) mode, the gradient of \(\theta\) approximates a straight line, indicating that the discrete wave vectors are equally spaced. In a closed universe, these wave vectors must be integers, constraining the gradient to be \(N\frac{\pi}{2}\) or \(\frac{1}{N}\frac{\pi}{2}\) (where \(N\in\mathbb{N}\))..
     }
     \label{fig:theta_kint}
\end{figure}

\begin{figure}
     \includegraphics{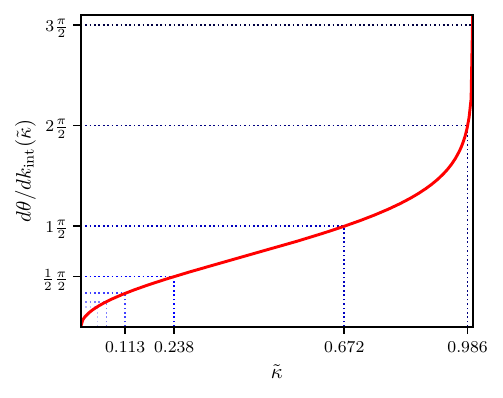}
     \caption{The gradient of the phase of perturbations \(\theta\), as a function of \(\tilde\kappa\). The gradient constraint determines the preferred curvature values.}
     \label{fig:Slope_kc}
\end{figure}

In \cref{fig:theta_diffK} \(\theta(\tilde{k})\) approaches a straight line for high-\(k\) modes due to perturbations oscillating as \(\sim \cos(\tilde{k}\eta/\sqrt{3})\).
This implies that the allowed wave vectors are equally spaced for high-\(k\) modes. By adjusting \(\tilde{\kappa}\), we can align the allowed wavevectors to be integers. Note that the unit of \(\tilde{k}\) differs from that of the integer wave vectors \(k_{\text{int}}\). The transformation between them can be expressed as: \(\tilde{k}\equiv  \sqrt{\frac{2\tilde{\kappa}}{3}} \sqrt{k_{\text{int}}(k_{\text{int}}+2) }  \approx \sqrt{\frac{2\tilde{\kappa}}{3}} k_{\text{int}}\) in high-\(k\) modes. Consequently, the actual gradient is given by:
\begin{equation}
\frac{d\theta}{dk_{\text{int}}}(\tilde{\kappa},\tilde{k}\gg 1,a\rightarrow \infty) \approx \sqrt{\frac{2\tilde{\kappa}}{3}}\frac{\eta_{\infty}(\tilde{\kappa})}{\sqrt{3}} \overset{!}{=} N^{\pm 1}\frac{\pi}{2},
\end{equation}
where to match the integer wave vectors, the gradient should be \(N\frac{\pi}{2}\) or \(\frac{1}{N}\frac{\pi}{2}\) (where \(N\in\mathbb{N}\)), corresponding to either sparser discrete \(\theta\) or sparser \(k_{\text{int}}\) (see \cref{fig:theta_kint}). Since the gradient depends on \(\tilde{\kappa}\), this constrains the allowed curvature values. Some allowed \(\tilde{\kappa}\) values are shown in \cref{fig:Slope_kc}. The gradient increases with \(\tilde{\kappa}\) and approaches infinity as \(\tilde{\kappa}\) approaches unity, the maximum allowed curvature value. Beyond this value, the universe would collapse, resulting in a crunch.

\begin{figure}
     \includegraphics{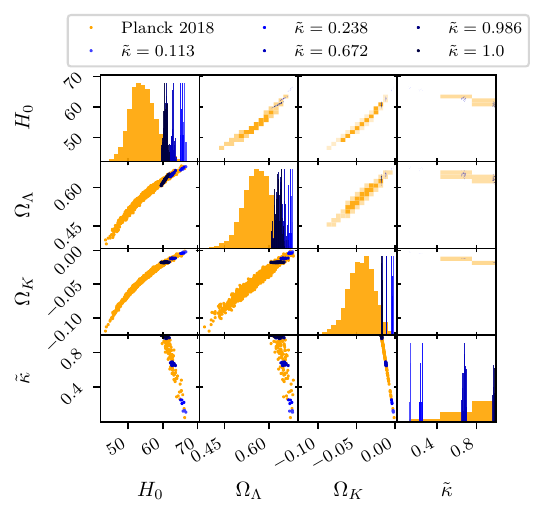}
     \caption{Comparison of predicted curvature values with \texttt{Planck 2018} data (\texttt{base\_omegak\_plikHM\_TTTEEE\_lowl\_lowE} from~\cite{2020A&A...641A...6P}). Predicted curvature values from \cref{fig:Slope_kc} are shown.}
     \label{fig:Planck_kappa}
\end{figure}

The predicted values of \(\tilde{\kappa}\) are then transformed to the current curvature density \(\Omega_K=-2\tilde{\kappa}\sqrt{\Omega_\Lambda\Omega_r}\) and compared with the \texttt{Planck 2018} data (\cref{fig:Planck_kappa}). The \(\tilde{\kappa}\), for unit gradient, corresponds to \(\Omega_K\sim -0.009\). The maximum curvature, \(\tilde{\kappa}=1\), corresponds to \(\Omega_K\sim -0.014\). Although this is closer to flat than the observational value of \(-0.044^{+0.018}_{-0.015}\)~\cite{2021PhRvD.103d1301H,2020NatAs...4..196D,2020A&A...641A...6P}, incorporating matter into our model would likely yield more precise predictions. Despite its simplicity, our model recovers curvature values consistent with observational data. 

\section{Conclusion} 
In this letter, we extended the work of \citet{2022PhRvD.105h3514L} and \citet{2021arXiv210906204B} by solving the first-order perturbation equation to derive analytic solutions. These solutions remain physical if they exhibit symmetry or antisymmetry at the FCB. This temporal boundary condition imposes a discrete spectrum on the wave vectors, which conflicts with the integer wave vectors derived from the spatial compactness of closed universes. Since both sets of discrete \(k\) values exhibit equal spacing in the high \(k\) regime, we can reconcile them by constraining the spatial curvature to specific values.. 

To refine the predictions of this allowed curvature spectrum further, we plan to solve the perturbations numerically in a more detailed universe model that includes matter and radiation anisotropy. The resulting solutions will be used to generate the CMB power spectra, which will then be compared with observational data, following the methodologies of~\cite{2022PhRvD.105h3515B} and~\cite{2022PhRvD.105l3508P}.

\begin{acknowledgments}
\section{acknowledgments}
We extend our gratitude to Latham Boyle for his helpful feedback on the calculations and logic presented in this paper. WND was supported by a Cambridge Trusts Taiwanese Scholarship. WJH was supported by a Royal Society University Research Fellowship. 

\end{acknowledgments}
\vfill
\pagebreak

% \appendix

% \section{Perturbation solution in Heun function form}

% \begin{equation}
% \begin{aligned}
%     &\Phi(a) = 
%     \left(1 + 2(1 - 2 \tilde\kappa^2) a^4 + a^8\right)^{1/8} \left(\frac{i a^2}{\alpha} - 1\right)^{-\gamma} (i \alpha a^2 + 1)^{\gamma} \\
%     &\times \exp\left(-\frac{1}{4}\tanh^{-1}\left(\frac{2a^2\tilde\kappa}{1+a^4}\right)+\frac{i +\tilde\kappa}{2\sqrt{\tilde\kappa^2-1}} \tanh^{-1} \left(\frac{-a^2+\tilde\kappa}{\sqrt{\tilde\kappa^2-1}}\right) \right)\\
%     &\times \text{HeunG}\left(-\frac{1}{\alpha^2}, -\frac{i \tilde{k}^2-i8\tilde\kappa - 5 \alpha}{4 \alpha}, 1, \frac{5}{2}, \frac{5}{2}, \frac{1}{2}, \frac{i a^2}{\alpha}\right),
% \end{aligned}
% \end{equation}

% where \(\alpha\) and \(\gamma\) are given by:
% \begin{equation*}
%     \alpha = i \tilde\kappa + \sqrt{1 - \tilde\kappa^2}, \quad \gamma = \frac{\alpha(\alpha - 1)}{2 \alpha^2 + 2}.
% \end{equation*}

\let\oldbibitem\bibitem
\renewcommand{\bibitem}{%
    \renewcommand{\doi}[1]{doi: ##1}% Override \doi
    \let\bibitem\oldbibitem% Restore \bibitem
    \oldbibitem% Call old \bibitem
}

\bibliography{Reference}% Produces the bibliography via BibTeX.

\begin{thebibliography}{29}
\providecommand{\natexlab}[1]{#1}
\providecommand{\url}[1]{\texttt{#1}}
\expandafter\ifx\csname urlstyle\endcsname\relax
  \providecommand{\doi}[1]{doi: #1}\else
  \providecommand{\doi}{doi: \begingroup \urlstyle{rm}\Url}\fi

\bibitem[{Boyle} et~al.(2018){Boyle}, {Finn}, and {Turok}]{2018PhRvL.121y1301B}
Latham {Boyle}, Kieran {Finn}, and Neil {Turok}.
\newblock {C P T -Symmetric Universe}.
\newblock \emph{\prl}, 121\penalty0 (25):\penalty0 251301, December 2018.
\newblock \doi{10.1103/PhysRevLett.121.251301}.

\bibitem[{Boyle} et~al.(2022{\natexlab{a}}){Boyle}, {Finn}, and
  {Turok}]{2022AnPhy.43868767B}
Latham {Boyle}, Kieran {Finn}, and Neil {Turok}.
\newblock {The Big Bang, CPT, and neutrino dark matter}.
\newblock \emph{Annals of Physics}, 438:\penalty0 168767, March
  2022{\natexlab{a}}.
\newblock \doi{10.1016/j.aop.2022.168767}.

\bibitem[{Lasenby} et~al.(2022){Lasenby}, {Handley}, {Bartlett}, and
  {Negreanu}]{2022PhRvD.105h3514L}
A.~N. {Lasenby}, W.~J. {Handley}, D.~J. {Bartlett}, and C.~S. {Negreanu}.
\newblock {Perturbations and the future conformal boundary}.
\newblock \emph{\prd}, 105\penalty0 (8):\penalty0 083514, April 2022.
\newblock \doi{10.1103/PhysRevD.105.083514}.

\bibitem[{Planck Collaboration}(2020)]{2020A&A...641A...6P}
{Planck Collaboration}.
\newblock {Planck 2018 results. VI. Cosmological parameters}.
\newblock \emph{\aap}, 641:\penalty0 A6, September 2020.
\newblock \doi{10.1051/0004-6361/201833910}.

\bibitem[{Peskin} and {Schroeder}(1995)]{1995iqft.book.....P}
Michael~E. {Peskin} and Daniel~V. {Schroeder}.
\newblock \emph{{An Introduction to Quantum Field Theory}}.
\newblock 1995.

\bibitem[{Dicke}(1972)]{1972grun.book.....D}
R.~H. {Dicke}.
\newblock \emph{{Gravitation and the universe.}}
\newblock 1972.

\bibitem[{Carrigan} and {Trower}(1983)]{1983Natur.305..673C}
Jr. {Carrigan}, R.~A. and W.~P. {Trower}.
\newblock {Magnetic monopoles}.
\newblock \emph{\nat}, 305\penalty0 (5936):\penalty0 673--678, October 1983.
\newblock \doi{10.1038/305673a0}.

\bibitem[{Weinberg}(1989)]{1989RvMP...61....1W}
Steven {Weinberg}.
\newblock {The cosmological constant problem}.
\newblock \emph{Reviews of Modern Physics}, 61\penalty0 (1):\penalty0 1--23,
  January 1989.
\newblock \doi{10.1103/RevModPhys.61.1}.

\bibitem[{Duff}(1994)]{1994CQGra..11.1387D}
M.~J. {Duff}.
\newblock {REVIEW ARTICLE: Twenty years of the Weyl anomaly}.
\newblock \emph{Classical and Quantum Gravity}, 11\penalty0 (6):\penalty0
  1387--1403, June 1994.
\newblock \doi{10.1088/0264-9381/11/6/004}.

\bibitem[{'t Hooft}(1976)]{1976PhRvL..37....8T}
G.~{'t Hooft}.
\newblock {Symmetry Breaking through Bell-Jackiw Anomalies}.
\newblock \emph{\prl}, 37\penalty0 (1):\penalty0 8--11, July 1976.
\newblock \doi{10.1103/PhysRevLett.37.8}.

\bibitem[{Guth}(1981)]{1981PhRvD..23..347G}
Alan~H. {Guth}.
\newblock {Inflationary universe: A possible solution to the horizon and
  flatness problems}.
\newblock \emph{\prd}, 23\penalty0 (2):\penalty0 347--356, January 1981.
\newblock \doi{10.1103/PhysRevD.23.347}.

\bibitem[{Profumo} et~al.(2019){Profumo}, {Giani}, and
  {Piattella}]{2019Univ....5..213P}
Stefano {Profumo}, Leonardo {Giani}, and Oliver~F. {Piattella}.
\newblock {An Introduction to Particle Dark Matter}.
\newblock \emph{Universe}, 5\penalty0 (10):\penalty0 213, October 2019.
\newblock \doi{10.3390/universe5100213}.

\bibitem[{Peccei} and {Quinn}(1977)]{1977PhRvL..38.1440P}
R.~D. {Peccei} and Helen~R. {Quinn}.
\newblock {CP conservation in the presence of pseudoparticles}.
\newblock \emph{\prl}, 38\penalty0 (25):\penalty0 1440--1443, June 1977.
\newblock \doi{10.1103/PhysRevLett.38.1440}.

\bibitem[{Clifton} et~al.(2012){Clifton}, {Ferreira}, {Padilla}, and
  {Skordis}]{2012PhR...513....1C}
Timothy {Clifton}, Pedro~G. {Ferreira}, Antonio {Padilla}, and Constantinos
  {Skordis}.
\newblock {Modified gravity and cosmology}.
\newblock \emph{\physrep}, 513\penalty0 (1):\penalty0 1--189, March 2012.
\newblock \doi{10.1016/j.physrep.2012.01.001}.

\bibitem[{Farren} et~al.(2024){Farren}, {Krolewski}, {MacCrann}, {Ferraro},
  {Abril-Cabezas}, {An}, {Atkins}, {Battaglia}, {Bond}, {Calabrese}, {Choi},
  {Darwish}, {Devlin}, {Duivenvoorden}, {Dunkley}, {Hill}, {Hilton},
  {Huffenberger}, {Kim}, {Louis}, {Madhavacheril}, {Marques}, {McMahon},
  {Moodley}, {Page}, {Partridge}, {Qu}, {Schaan}, {Sehgal}, {Sherwin},
  {Sif{\'o}n}, {Staggs}, {Van Engelen}, {Vargas}, {Wenzl}, {White}, and
  {Wollack}]{2024ApJ...966..157F}
Gerrit~S. {Farren}, Alex {Krolewski}, Niall {MacCrann}, Simone {Ferraro}, Irene
  {Abril-Cabezas}, Rui {An}, Zachary {Atkins}, Nicholas {Battaglia}, J.~Richard
  {Bond}, Erminia {Calabrese}, Steve~K. {Choi}, Omar {Darwish}, Mark~J.
  {Devlin}, Adriaan~J. {Duivenvoorden}, Jo~{Dunkley}, J.~Colin {Hill}, Matt
  {Hilton}, Kevin~M. {Huffenberger}, Joshua {Kim}, Thibaut {Louis}, Mathew~S.
  {Madhavacheril}, Gabriela~A. {Marques}, Jeff {McMahon}, Kavilan {Moodley},
  Lyman~A. {Page}, Bruce {Partridge}, Frank~J. {Qu}, Emmanuel {Schaan}, Neelima
  {Sehgal}, Blake~D. {Sherwin}, Crist{\'o}bal {Sif{\'o}n}, Suzanne~T. {Staggs},
  Alexander {Van Engelen}, Cristian {Vargas}, Lukas {Wenzl}, Martin {White},
  and Edward~J. {Wollack}.
\newblock {The Atacama Cosmology Telescope: Cosmology from Cross-correlations
  of unWISE Galaxies and ACT DR6 CMB Lensing}.
\newblock \emph{\apj}, 966\penalty0 (2):\penalty0 157, May 2024.
\newblock \doi{10.3847/1538-4357/ad31a5}.

\bibitem[{Boyle} and {Turok}(2021{\natexlab{a}})]{2021arXiv211006258B}
Latham {Boyle} and Neil {Turok}.
\newblock {Cancelling the vacuum energy and Weyl anomaly in the standard model
  with dimension-zero scalar fields}.
\newblock \emph{arXiv e-prints}, art. arXiv:2110.06258, October
  2021{\natexlab{a}}.
\newblock \doi{10.48550/arXiv.2110.06258}.

\bibitem[{Boyle} et~al.(2022{\natexlab{b}}){Boyle}, {Teuscher}, and
  {Turok}]{2022arXiv220810396B}
Latham {Boyle}, Martin {Teuscher}, and Neil {Turok}.
\newblock {The Big Bang as a Mirror: a Solution of the Strong CP Problem}.
\newblock \emph{arXiv e-prints}, art. arXiv:2208.10396, August
  2022{\natexlab{b}}.
\newblock \doi{10.48550/arXiv.2208.10396}.

\bibitem[{Boyle} and {Turok}(2024)]{2024PhLB..84938442B}
Latham {Boyle} and Neil {Turok}.
\newblock {Thermodynamic solution of the homogeneity, isotropy and flatness
  puzzles (and a clue to the cosmological constant)}.
\newblock \emph{Physics Letters B}, 849:\penalty0 138442, February 2024.
\newblock \doi{10.1016/j.physletb.2024.138442}.

\bibitem[{Turok} and {Boyle}(2023)]{2023arXiv230200344T}
Neil {Turok} and Latham {Boyle}.
\newblock {A Minimal Explanation of the Primordial Cosmological Perturbations}.
\newblock \emph{arXiv e-prints}, art. arXiv:2302.00344, February 2023.
\newblock \doi{10.48550/arXiv.2302.00344}.

\bibitem[{Gibbons} and {Hawking}(1977)]{1977PhRvD..15.2752G}
G.~W. {Gibbons} and S.~W. {Hawking}.
\newblock {Action integrals and partition functions in quantum gravity}.
\newblock \emph{\prd}, 15\penalty0 (10):\penalty0 2752--2756, May 1977.
\newblock \doi{10.1103/PhysRevD.15.2752}.

\bibitem[{Turok} and {Boyle}(2024)]{2024PhLB..84938443T}
Neil {Turok} and Latham {Boyle}.
\newblock {Gravitational entropy and the flatness, homogeneity and isotropy
  puzzles}.
\newblock \emph{Physics Letters B}, 849:\penalty0 138443, February 2024.
\newblock \doi{10.1016/j.physletb.2024.138443}.

\bibitem[{Bartlett} et~al.(2022){Bartlett}, {Handley}, and
  {Lasenby}]{2022PhRvD.105h3515B}
D.~J. {Bartlett}, W.~J. {Handley}, and A.~N. {Lasenby}.
\newblock {Improved cosmological fits with quantized primordial power spectra}.
\newblock \emph{\prd}, 105\penalty0 (8):\penalty0 083515, April 2022.
\newblock \doi{10.1103/PhysRevD.105.083515}.

\bibitem[{Prathaban} and {Handley}(2022)]{2022PhRvD.105l3508P}
Metha {Prathaban} and Will {Handley}.
\newblock {Rescuing palindromic universes with improved recombination
  modeling}.
\newblock \emph{\prd}, 105\penalty0 (12):\penalty0 123508, June 2022.
\newblock \doi{10.1103/PhysRevD.105.123508}.

\bibitem[{Di Valentino} et~al.(2020){Di Valentino}, {Melchiorri}, and
  {Silk}]{2020NatAs...4..196D}
Eleonora {Di Valentino}, Alessandro {Melchiorri}, and Joseph {Silk}.
\newblock {Planck evidence for a closed Universe and a possible crisis for
  cosmology}.
\newblock \emph{Nature Astronomy}, 4:\penalty0 196--203, February 2020.
\newblock \doi{10.1038/s41550-019-0906-9}.

\bibitem[{Handley}(2021)]{2021PhRvD.103d1301H}
Will {Handley}.
\newblock {Curvature tension: Evidence for a closed universe}.
\newblock \emph{\prd}, 103\penalty0 (4):\penalty0 L041301, February 2021.
\newblock \doi{10.1103/PhysRevD.103.L041301}.

\bibitem[{Glanville} et~al.(2022){Glanville}, {Howlett}, and
  {Davis}]{2022MNRAS.517.3087G}
Aaron {Glanville}, Cullan {Howlett}, and Tamara {Davis}.
\newblock {Full-shape galaxy power spectra and the curvature tension}.
\newblock \emph{\mnras}, 517\penalty0 (2):\penalty0 3087--3100, December 2022.
\newblock \doi{10.1093/mnras/stac2891}.

\bibitem[{Efstathiou} and {Gratton}(2020)]{2020MNRAS.496L..91E}
George {Efstathiou} and Steven {Gratton}.
\newblock {The evidence for a spatially flat Universe}.
\newblock \emph{\mnras}, 496\penalty0 (1):\penalty0 L91--L95, July 2020.
\newblock \doi{10.1093/mnrasl/slaa093}.

\bibitem[{Boyle} and {Turok}(2021{\natexlab{b}})]{2021arXiv210906204B}
Latham {Boyle} and Neil {Turok}.
\newblock {Two-Sheeted Universe, Analyticity and the Arrow of Time}.
\newblock \emph{arXiv e-prints}, art. arXiv:2109.06204, September
  2021{\natexlab{b}}.
\newblock \doi{10.48550/arXiv.2109.06204}.

\bibitem[{Tram}(2017)]{2017CCoPh..22..852T}
Thomas {Tram}.
\newblock {Computation of Hyperspherical Bessel Functions}.
\newblock \emph{Communications in Computational Physics}, 22\penalty0
  (3):\penalty0 852--862, September 2017.
\newblock \doi{10.4208/cicp.OA-2016-0071}.

\end{thebibliography}

\end{document}